\documentclass[aps,prb,twocolumn,superscriptaddress,
showpacs,preprintnumbers,amsmath,amssymb
,floatfix
]{revtex4}

\usepackage{graphicx}
\usepackage{dcolumn}
\usepackage{bm}

\begin{document}

\title{Does Nature Allow Negative Refraction with Low Losses in 
Optical Region?}

\author{Mark I. Stockman}
\email{mstockman@gsu.edu}
\homepage{http://www.phy-astr.gsu.edu/stockman}
\affiliation{Department of Physics and Astronomy, Georgia State
University, Atlanta, GA 30303, USA
}

\date{\today}

\begin{abstract}
From  the fundamental requirement of causality, we derive a rigorous criterion of
negative refraction (left-handedness). This
criterion imposes the lower limits on the electric and magnetic
losses in the region of the negative refraction. If these losses are eliminated or
significantly reduced  by any means, including the compensation by active (gain)
media, then the negative refraction will disappear. This theory can be particularly 
useful in designing new left-handed materials: testing the expected 
polarizabilities of a medium against this criterion would check the compliance 
with the causality and verify the design feasibility.
\end{abstract}


\maketitle

There has been recently a significant attention devoted to the so called
left-handed materials (LHM), which are also called negative-refraction
media 
\cite{Pendry_Opt_Expr_2003_Introduction_to_Focus_Issue_on_LHM,%
Smith_PRL_2000_Left_Handed_Material,%
Sheldon_Schultz_et_al_Phys_Rev_Lett_84_4184_2000_Left_Handed_Material,%
Shelby_et_al_APL_2001_Left_Handed_Metamaterial,%
Houck_et_al_PRL_2003_LHM,%
Parazzoli_et_al_PRL_2003_LHM,%
Lagarkov_Kissel_Phys_Rev_Lett_92_077401_2004,%
Osgood_et_al_PRL_2005_Near_IR_LHM,%
Grigorenko_et_al_Nature_2005_Left_Handed_Metanaterial,%
Shalaev_et_al_OL_2005_LHM,Wegener_et_al_Science_2006_LHM,%
Leonhardt_Science_2006_EM_Cloacking,%
Pendry_Schurig_Smith_Science_2006_EM_Cloacking}.
In such materials, the directions
of energy transfer and wave-front propagation are opposite. This leads
to remarkable electromagnetic (optical) properties such as 
refraction at surfaces that is
described by a negative refraction index $n$. This, in turn, causes a flat
slab of a left-handed material with $n=-1$ to act as a ``perfect'' lens creating,
without reflections at the surfaces, a
non-distorted image. This is a so-called Veselago lens \cite{Veselago:1968}.
Moreover, such a lens can also build an image in the near field 
\cite{Pendry_Phys_Rev_Lett_85_3996_2000:Perfect_Lens}. Optical losses in
LHMs are detrimental to their performance. These
losses for LHMs in the near-infrared and visible region are significant 
\cite{Osgood_et_al_PRL_2005_Near_IR_LHM,%
Grigorenko_et_al_Nature_2005_Left_Handed_Metanaterial,%
Shalaev_et_al_OL_2005_LHM,Wegener_et_al_Science_2006_LHM}, which
drastically limits their usefulness. There have been proposals to
compensate these losses with optical gain 
\cite{Nezhad_Tetz_Fainman_Opt_Expr_2004_Gain_Compensated_Loss_of_SPPs,%
Popov_Shalaev_Opt_Lett_2006_Parametric_Loss_Compensation}, which
appears to be a way to resolve this problem. 
In this Letter, we show
that compensating the optical losses, which implies significantly
reducing the imaginary part of the dielectric permittivity $\varepsilon$
and magnetic permeability $\mu$, will necessarily change also the real
parts of these quantities in such a way that the negative refraction
disappears. This follows from the dispersion relations,
i.e., ultimately, from the fundamental principle of causality.

A proposal to add a gain medium to a metal nanosystem to create a
nanoplasmonic counterpart of laser (spaser) in the near-infrared and
visible spectral region has been introduced in Refs.\
\onlinecite{Bergman_Stockman:2003_PRL_spaser,Spaser_SPIE_Proceedings_2003}.
Earlier, a THz quantum cascade laser with surface-polariton resonator has
been created \cite{Colombelli_Capasso_et_al:2001_sp_laser}. A
possibility to compensate a small fraction of optical losses in
plasmonic propagation by gain as a first step toward creation a spaser
has been experimentally shown 
\cite{Seidel_Grafstroem_Eng_Phys_Rev_Lett_94_177401_2005}. Using optical
gain to compensate losses in the plasmonic ``perfect lens'' has been
proposed \cite{Ramakrishna_Pendry:2003_PRB_201101_Spaser_Superlense}.
The interest to the compensation of losses in the metal plasmonic
systems by the optical gain has attracted recently a great deal of
attention in conjunction with the formidable problem of creating LHMs 
in the near-infrared and visible spectrum with low losses 
\cite{Nezhad_Tetz_Fainman_Opt_Expr_2004_Gain_Compensated_Loss_of_SPPs,%
Popov_Shalaev_Opt_Lett_2006_Parametric_Loss_Compensation,%
Popov_Shalaev_Appl_Phys_B_2006_SHG_and_Parametric_Gain}. This
compensation of losses in LHMs by gain appears to be very attractive since
the existing implementations of the LHMs in the optical region suffer
from large optical losses: $|\mathrm{Im}\,k|\sim |\mathrm{Re}\,k|$,
where $k$ is the wave vector, so a wave propagates just a few periods in
such a medium before extinction reduces its intensity several times
\cite{Osgood_et_al_PRL_2005_Near_IR_LHM,%
Grigorenko_et_al_Nature_2005_Left_Handed_Metanaterial,%
Shalaev_et_al_OL_2005_LHM,Wegener_et_al_Science_2006_LHM}. Apart from
the active approach based on the gain media, there is also a possibility
to use different materials or to nanostructure a medium to lower the
optical losses. 

However, it is impossible to reduce the optical losses
without affecting the real part of the dielectric and magnetic responses
because of the requirements of causality leading to the familiar
Kramers-Kronig dispersion relations (see, e.g., Ref.\
\onlinecite{Landau_Lifshitz_Electrodynamics_Continuous:1984}). In this
Letter, we derive similar dispersion relations for the squared
refractive index. Using them we show that a significant reduction in the
optical losses at and near the observation frequency will necessarily
eliminate the negative refraction. 

The Kramers-Kronig relations follow from the causality of the dielectric
response function in the temporal domain. Then
one can prove that in the frequency domain permittivity
$\varepsilon(\omega)$ does not have singularities in the upper
half-plane of the complex variable $\omega$. From this and the limit
$\varepsilon(\omega)\to 1$ for $\omega\to \infty$, one derives the
conventional Kramers-Kronig dispersion relation for the dielectric
function. For the same causality reason, magnetic permeability
$\mu(\omega)$ does not have singularities in the upper half-plane of
complex $\omega$. Since also $\mu(\omega)\to 1$ for $\omega\to \infty$,
permeability $\mu(\omega)$ satisfies a similar dispersion relation. Note
the requirement of the response linearity is essential: nonlinear and
saturated polarizabilities generally do not satisfy the Kramers-Kronig
relations. \cite{Boyd_Nonlinear_Optics_Second_Edition} 
We will below consider systems including gain media; in
those cases we assume that the optical reponses to the {\it signal}
(observed) radiation are linear. This of course requires the signal to
be weak enough to ensure the linearity of the responses to it and the
applicability of the Kramers-Kronig relations.

We consider a material to be an effective medium characterized by macroscopic
permittivity $\varepsilon(\omega)$ and permeability $\mu(\omega)$.
The squared complex refraction index
$n^2(\omega)=\varepsilon(\omega)\mu(\omega)$ has exactly the same
analytical properties as $\varepsilon(\omega)$ and $\mu(\omega)$ separately:
$n^2(\omega)$ does not have singularities in the upper half plane of
complex $\omega$ and $n^2(\omega)\to 1$ for $\omega\to \infty$. 
Therefore, absolutely similar to the derivation of
the Kramers-Kronig relations for the permittivity or permeability (see,
e.g., Ref.\
\onlinecite{Landau_Lifshitz_Electrodynamics_Continuous:1984}), we obtain
a dispersion relation for $n^2(\omega)$,
\begin{equation}
\mathrm{Re}\,n^2(\omega)=1+\frac{2}{\pi}
\mathcal{P}\int_0^\infty\frac{\mathrm{Im}\,n^2(\omega_1)}
{\omega_1^2-\omega^2}\omega_1\,\mathrm d\omega_1~,
\label{dis_rel}
\end{equation}
where $\mathcal{P}$ denotes the principal value of an integral. 

Note that in contrast to $n^2(\omega)$, refractive index
$n(\omega)=\sqrt{n^2}$ may possess singularities
in the upper half plane and thus is generally not causal; this is true,
in particular, when optical gain is present
\cite{Skaar_PRE_73_026605_2006_Causality_n_with_Gain}. The refractive
index $n$ {\it per se} does not enter the Maxwell equations; it is not a
susceptibility, and it does not have to obey the causality, while $n^2$
does. This theory is based on $n^2$, not $n$; the non-causality of $n$
is irrelevant for its purposes.

Now we assume that at the observation
frequency $\omega$ the material is transparent (e.g., the losses are
compensated by gain), i.e.,
$\mathrm{Im}\,n^2(\omega)=0$ with any required accuracy. Then
the principal value in the right-hand side of Eq.\ (\ref{dis_rel}) can
be omitted. Multiplying both sides of this equation by
$\omega^2$ and differentiating over $\omega$ (one can differentiate
under the integral over $\omega$ as a parameter, because the point
$\omega_1=\omega$ is not singular anymore), we obtain
\begin{equation}
\frac{\partial \omega^2 \left[\mathrm{Re}\,n^2(\omega)-1\right]}
{\partial\omega}=\frac{4\omega}{\pi}
\int_0^\infty\frac{\mathrm{Im}\,n^2(\omega_1)}
{(\omega_1^2-\omega^2)^2}\,\omega_1^3\,\mathrm d\omega_1~.
\label{dis_rel_1}
\end{equation}

The left-hand side of this equation can be expressed in terms of the
phase velocity $\mathbf v_p=(\mathbf k/k)\, \omega/k$, where wave vector
$k=\omega n(\omega)/c$ and $c$ is speed of light, and group velocity
$\mathbf v_g=(\mathbf k/k)\, \partial \omega/\partial k$. In this
way, we obtain
\begin{eqnarray}
&&\displaystyle\frac{1}{\mathbf v_p \mathbf v_g}-\frac{1}{c^2}=
\nonumber\\
&&\displaystyle\frac{2}{\pi c^2} \int_0^\infty
\frac{\varepsilon^{\prime\prime}(\omega_1) \mu^\prime(\omega_1) +
\mu^{\prime\prime}(\omega_1) \varepsilon^\prime(\omega_1)}
{(\omega_1^2-\omega^2)^2}\,\omega_1^3\,\mathrm d\omega_1~,
\label{v_p_v_g}
\end{eqnarray}
where $\varepsilon^\prime=\mathrm{Re}\, \varepsilon$,
$\varepsilon^{\prime\prime}=\mathrm{Im}\,\varepsilon$ and, similarly, 
$\mu^\prime=\mathrm{Re}\, \mu$, $\mu^{\prime\prime}=\mathrm{Im}\,\mu$;
$\mathrm{Im}\,n^2(\omega)=
\varepsilon^{\prime\prime}(\omega) \mu^\prime(\omega) +
\mu^{\prime\prime}(\omega) \varepsilon^\prime(\omega)$.

In the case of the negative refraction, the directions of the phase and
energy propagation are opposite, therefore $\mathbf v_p \mathbf v_g<0$.
Consequently, we obtain from Eq.\ (\ref{v_p_v_g}) a rigorous criterion of
the negative refraction with no (or low) 
loss at the observation frequency $\omega$ as
\begin{equation}
\frac{2}{\pi} \int_0^\infty
\frac{\varepsilon^{\prime\prime}(\omega_1) \mu^\prime(\omega_1) +
\mu^{\prime\prime}(\omega_1) \varepsilon^\prime(\omega_1)}
{(\omega_1^2-\omega^2)^2}\,\omega_1^3\,\mathrm d\omega_1
\le -1~.
\label{criterion}
\end{equation}
This criterion directly imposes the lower 
bounds on the dielectric losses [$\varepsilon^{\prime\prime}(\omega_1)>0$], 
overlapping with the
magnetic plasmonic behavior [$\mu^\prime(\omega_1)<0$] and 
the magnetic losses [$\mu^{\prime\prime}(\omega_1)>0$] overlapping
with the electric plasmonic behavior [$\varepsilon^\prime(\omega_1)<0$].
The denominator $(\omega_1^2-\omega^2)^2$ makes the
integral to converge for $|\omega_1-\omega|$ large; it would have
diverged at  $|\omega_1-\omega|\to 0$ if the integrand did not vanish at
that point. Thus, the major
contribution to Eq.\ (\ref{criterion})
comes from the lossy, overlapping electric and
magnetic resonances close to observation
frequency $\omega$.

The stability of the system 
requires that no net gains are
present at any frequency, i.e., $\varepsilon^{\prime\prime}(\omega)\ge 0$ and
$\mu^{\prime\prime}(\omega)\ge 0$ everywhere. 
\cite{Landau_Lifshitz_Electrodynamics_Continuous:1984}
There is a known condition of negative refraction
\cite{Depine_Lakhtakia_MWOTL_2004_New_Condition_for_Negative_Phase_Velocity}
$\mathrm{Im}\,n^2(\omega)<0$. This condition is always satisfied in
the region of left-handedness where $\varepsilon^\prime(\omega)<0$ and
$\mu^\prime(\omega)<0$. Thus, this condition is trivial: in contrast to
Eq.\ (\ref{criterion}), it does not impose a lower limit on the losses.

In the absence of magnetic resonances, in the optical region
$\mu^\prime=1$ and $\mu^{\prime\prime}=0$. Then it is
obvious than the integral in the left-hand side of Eq.\
(\ref{criterion}) is strictly positive and this criterion is not
satisfied, i.e., the negative refraction is absent. In the presence of a
magnetic resonance, in a part of its region $\mu^\prime<0$ and
$\mu^{\prime\prime}>0$; thus the criterion (\ref{criterion}) can, in
principle, be satisfied. However, this requires non-zero losses: 
$\mu^{\prime\prime}>0$ and/or $\varepsilon^{\prime\prime}>0$. 

To satisfy the transparency requirement at the observation frequency,
$\mathrm{Im}\,n^2(\omega)=0$, one may attempt to add a gain
to exactly cancel out the losses at this frequency 
\cite{Popov_Shalaev_Opt_Lett_2006_Parametric_Loss_Compensation,%
Popov_Shalaev_Appl_Phys_B_2006_SHG_and_Parametric_Gain}. 
Is it possible from
the positions of causality? Because the loss should nowhere be negative,
it is obvious that it must have the zero minimum at frequency $\omega$. 
The corresponding resonant contribution to the permittivity
close to resonance frequency $\omega_r$ has the form
\begin{equation}
\varepsilon_r(\omega_1)\propto\left[\omega_1-\omega_r+%
i\frac{1}{2}(\omega_1-\omega)^2\frac{\partial^2 \gamma(\omega)}%
{\partial\omega^2}\right]^{-1}~.
\label{epsilon_r}
\end{equation}
Here $\gamma(\omega)$ is the relaxation rate that depends on frequency due to
the loss compensation. At the observation frequency this loss is
completely compensated, $\gamma(\omega)=0$, and it has a minimum:
$\partial \gamma(\omega)/\partial \omega=0$ and 
$\partial^2 \gamma(\omega)/\partial \omega^2>0$.
However, it follows from this equation that $\varepsilon_r(\omega_1)$
has an extra pole at a complex frequency
\begin{equation}
\omega_1\approx \omega+2i
\left(\left.{\partial^2 \gamma(\omega)}\right/{\partial \omega^2}\right)^{-1}~.
\label{extra_pole}
\end{equation}
This pole is situated in the {\it upper} half plane, which {\it violates
causality}. Because the form of Eq.\ (\ref{epsilon_r}) is rather general
close to the resonance, we conclude that in this manner it is impossible 
to compensate the losses.

It is still possible that both the magnetic resonance and electric
plasmonic behavior are present, but their losses are
compensated by an active-medium gain. However, such compensation must
take place not only at the observation frequency $\omega$, but for the
{\it entire} region of such resonances assuming their homogeneous
nature. This means that in Eq.\ (\ref{criterion}) whenever
$\mu^\prime(\omega_1)<0$, we have $\mu^{\prime\prime}(\omega_1)=0$ and
$\varepsilon^{\prime\prime}(\omega_1)=0$. However, in this case the
contribution of this region to the integral in Eq.\ (\ref{criterion})
vanishes, and the contribution of the region of normal optical magnetic
behavior ($\mu=1$) is always positive. Consequently, the
negative-refraction criterion is violated, which implies the absence of
the negative refraction. 

To obtain the negative refraction, the losses in the magnetic resonance
region not only should be present, but they should be significant not
only to overcome the positive contribution of the non-resonant region to
the integral in Eq.\ (\ref{criterion}), but actually
to make it less than $-1$. Thus, significantly reducing by any means,
passive or active (by gain), the losses of the negative-refraction
resonances will necessarily eliminate this negative refraction itself.
Fundamentally, this stems from the fact that the imaginary part and real
part of the squared index of refraction are not independent but must
satisfy the requirements imposed by the principle of causality.

One has to explore also a possibility to satisfy the criterion
(\ref{criterion}) with low losses at the working frequency $\omega$ by
having a left-handed resonance somewhere else at some resonance
frequency $\omega_r$ remote from $\omega$ to satisfy Eq.\
(\ref{criterion}). The contribution of such a
remote resonance to the integral in Eq.\ (\ref{criterion}) can be
approximated as
\begin{equation}
\frac{2}{\pi}\frac{\omega_r^3}{(\omega_r^2-\omega^2)^2}\,
\mathrm{Im} \int_{-\infty}^\infty n_r^2(\omega_1)\,d\omega_1~.
\label{resonance}
\end{equation}
Here $n_r^2(\omega_1)$ is the resonant contribution to the squared
index. It is assumed that it decreases rapidly enough when $|\omega_1-
\omega_r|\to\infty$, which is the expression of its resonant behavior.
In this case, it is possible to extend the integral in this equation
over the entire region, as indicated. As required
by the causality, $n_r^2(\omega_1)$ does not have any singularities in
the upper half plane of $\omega_1$. This integral can be closed by an
infinite arc in the upper half-plane, which gives the zero result due to
this absence of the singularities there. Hence, the distant resonances do not
contribute to the negative-refraction criterion (\ref{criterion}). This
completes the proof that zero (or, very low) losses at and near the
observation frequency are incompatible with the negative refraction.

We point out that in reality these losses do not have to be zero
to eliminate the negative refraction. If they are
merely much smaller that the losses in the adjacent regions that result
in the positive contribution to the integral in criterion
(\ref{criterion}), then the negative refraction will
be absent. In the microwave region, these losses can actually be quite
small, but not so in the optical region.

Simple, exactly solvable, and convincing illustrations of the above
theory are provided by the negative refraction of surface plasmon
polaritons (SPPs) in films with nanoscale thickness. Note that it
is a two-dimensional refraction but our consideration is based on the
principle of causality and is general, applicable to refraction
in spaces of arbitrary dimensions. We emphasize the the examples to
follow do not provide a proof but serve merely as
illustrations of the above-given proof.

Consider a flat layer with nanoscale
thickness $d$ made of a material with dielectric permittivity
$\varepsilon_2$ embedded between two half-spaces of materials with
permittivities $\varepsilon_1$ and $\varepsilon_3$. The dispersion
relation, i.e., wave vector $k$ as a function of $\omega$ or {\it vice
versa}, of the waves (SPPs) bound to the nanolayer can be found from an
exact, analytical transcendental equation
\begin{equation}
\tanh{(\omega\, d\,\varepsilon_2\, u_2/c)}=
-\left.{u_2(u_1+u_2)}\right/({u_1u_3+u_2^2})~,
\label{char_eq_1_tanh}
\end{equation}
where $u_i=\varepsilon_i^{-1}\sqrt{\left(kc/\omega\right)^2-\varepsilon_i}$.

As the first example, we mention a semi-infinite metal (silver)
covered with a nanolayer of dielectric with a half-space of another
dielectric covering it.
\cite{Karalis_Lidorikis_Ibanescu_Joannopoulos_Soljacic_%
Phys_Rev_Lett_95_063901_2005_Plasmonic_Waveguiding,%
Stockman_Nano_Lett_2006_Slow_Propagation_Total_External_Reflection}
This system possesses an extended spectral region of negative
refraction
\cite{Stockman_Nano_Lett_2006_Slow_Propagation_Total_External_Reflection}; 
however, in this region the SPP losses are so high that the
propagation is actually absent, in
accord with the above-presented theory.

\begin{figure}
\centering
\includegraphics[width=.45\textwidth]
{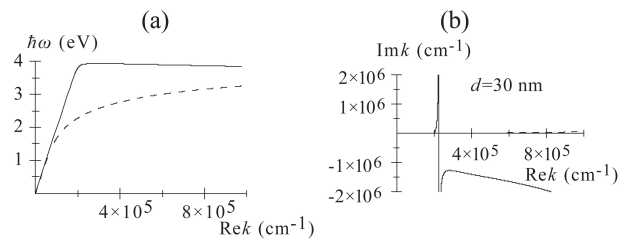}
\caption{\label{Thin_metal_layer_dispersion_relation.eps}
Dispersion relations for thin silver film in vacuum. The symmetric and
antisymmetric modes are displayed with solid and dashed lines,
respectively. (a) Real part of
dispersion relation: frequency $\omega$ as a function of $\mathrm{Re}k$.
(b) Imaginary part of the dispersion relation:
dependence of $\mathrm{Im}\,k$ on $\mathrm{re}\,k$. Thickness of the silver
film is $d=30$ nm.
}
\end{figure}

Another exactly solvable example of negative refraction also described
by Eq.\ (\ref{char_eq_1_tanh}) is given by SPPs in a metal film of a
nanoscopic thickness embedded in a dielectric. There are two
metal-dielectric interfaces and, correspondingly, two modes of SPPs in
this system. Because there is symmetry with respect to the reflection in
the middle plain, these SPP modes are classified according to their
magnetic-field parity: symmetric and antisymmetric. As an example, we
consider a silver film with thickness $d\sim 10$ nm in vacuum. The
corresponding dispersion relations are shown in Fig.\
\ref{Thin_metal_layer_dispersion_relation.eps}. As we see from panel
(a), the symmetric SPPs have regions of both the positive refraction
($\mathrm{Re}\,k<2\times10^5 ~\mathrm{cm^{-1}}$) and negative refraction
($\mathrm{Re}\,k>2\times10^5 ~\mathrm{cm^{-1}}$), while the
antisymmetric SPPs possess only the positive refraction. The optical
losses are shown in Fig.\
\ref{Thin_metal_layer_dispersion_relation.eps}(b). For most of the
positive-refraction region of the symmetric SPPs and in the entire
spectral range of the antisymmetric SPPs, these losses are relatively
very small: $\mathrm{Im}\,k\ll\mathrm{Re}\,k$. However, for the
symmetric SPPs (the solid line) close to the
negative-refraction region, the losses dramatically increase by orders
of magnitude. Inside the negative-refraction region, they are extremely
high, $|\mathrm{Im}\,k|\gtrsim |\mathrm{Re}\,k|$, so the propagation is
overdamped and actually absent,
in the full agreement with the conclusions of the present
theory. \footnote{In this region
$\mathrm{Im}\,k<0$ describing the propagation of energy opposite to the
wave vector.} 

\begin{figure}
\centering
\includegraphics[width=.45\textwidth]
{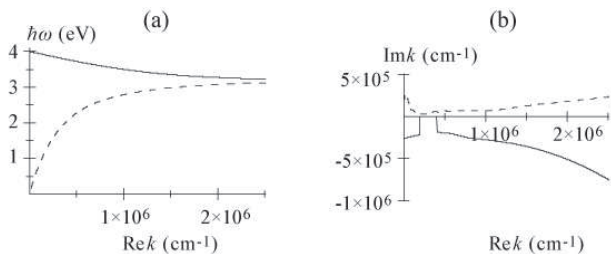}
\caption{\label{Re_Im_k_Gap.eps}
(a) For a thin ($d=10$ nm)
dielectric layer with $\varepsilon_d=3$ embedded in thick
metal (silver), dispersion relation of SPPs 
displayed of  as dependence of frequency
$\hbar\omega$ on the real part $\mathrm{Re}\,k$ of wave vector. (b) For
the same system, dependence of the wavevector imaginary part
$\mathrm{Im}\,k$ on its real part $\mathrm{Re}\,k$. For both panels, the 
solid lines
pertain to the antisymmetric SPP mode, and the dashed lines denote the
symmetric SPP mode.
}
\end{figure}

Yet another system that supports the negative-refraction SPPs is 
a dielectric nanolayer embedded in a metal
\cite{Shin_Zhang_PRL_96_073907_2006_SPP_Negative_Refraction}. This
system is also symmetric and possesses two metal-dielectric interfaces.
Therefore it supports two branches of SPPs that are characterized by
parity. The corresponding dispersion relations are displayed in
Fig.\ \ref{Re_Im_k_Gap.eps}. The real parts of these dispersion relations
as functions $\omega(\mathrm{Re}\,k)$ for these two types of modes 
are shown in panel(a). From it we see that in the entire spectral region
the symmetric SPPs (dashed line) have normal, positive
refraction ($v_g>0$), while the antisymmetric SPPs (solid line) are
negative-refracting ($v_g<0$). The corresponding losses 
are displayed in Fig.\
\ref{Re_Im_k_Gap.eps}(b). We note that the losses of the
positive-refraction, symmetric mode (dashed line) are relatively small in
the entire region, $\mathrm{Im}\,k\ll\mathrm{Re}\,k$. In a sharp
contrast, for the antisymmetric, negative-refraction mode (solid line),
the losses for small wave vectors are very high, 
$|\mathrm{Im}\,k|\gtrsim\mathrm{Re}\,k$, so the wave
propagates through only a few periods before it dissipates. 
The apparent discontinuity of the
corresponding curve at small momenta is due to the failure of the
numerical procedure to find a root of characteristic equation
(\ref{char_eq_1_tanh}), which a consequence of the fact that 
a good, propagating SPP mode in this spectral region 
does not exist.

To conclude, from the fundamental principle of causality, we have
derived a dispersion relation (\ref{dis_rel}) for the squared refraction
index. From it, assuming a low loss at the observation frequency,
we have derived a criterion (\ref{criterion}) of the negative
refraction. We have shown that the low loss at and near
the observation frequency is incompatible with the existence of the
negative refraction \footnote{The low loss is understood as
significantly lower than in the spectral regions adjacent to the
magnetic resonance, or the loss that would have be present if the gain
medium compensation were not introduced.}. While at the THz region the
losses may not be significant, they are very large in the optical
region. The loss
compensation or significant reduction will necessarily lead to the
disappearance of the negative refraction itself due to the
dispersion relation dictated by the causality.

This work was supported by grants from the Chemical Sciences, Biosciences and
Geosciences Division of the Office of Basic Energy Sciences, Office of
Science, U.S. Department of Energy, a grant CHE-0507147 from NSF, 
and a grant from the US-Israel BSF. 



\end{document}